\documentstyle[12pt]{article}
\def\gev{\; {\rm GeV}}

\def\mbar{\overline{m}}
\def\veff{V_{eff}(\phi,T)}
\def\v1t{V^{(1)}[m^2(\phi),T]}

\def\ov{\overline}
\def\simlt{\mathrel{\lower2.5pt\vbox{\lineskip=0pt\baselineskip=0pt
           \hbox{$<$}\hbox{$\sim$}}}}
\def\simgt{\mathrel{\lower2.5pt\vbox{\lineskip=0pt\baselineskip=0pt
           \hbox{$>$}\hbox{$\sim$}}}}
\def\to{\rightarrow}
\newcommand{\be}{\begin{equation}}
\newcommand{\ee}{\end{equation}}
\newcommand{\bea}{\begin{eqnarray}}
\newcommand{\eea}{\end{eqnarray}}
\newcommand{\bean}{\begin{eqnarray*}}
\newcommand{\eean}{\end{eqnarray*}}
\begin{document}
\begin{titlepage}
\vspace*{-1cm}
\noindent
\phantom{DRAFT}
\hfill{CERN-TH.6451/92}
\\
\phantom{bla}
\hfill{IEM-FT-56/92}
\vskip 2.5cm
\begin{center}
{\Large\bf On the phase transition in the scalar theory}
\end{center}
\vskip 1.5cm
\begin{center}
{\large J. R. Espinosa}\footnote{Also at Instituto de Estructura de la
Materia, Madrid, Spain.}$^,$\footnote{Supported by a grant of Fundaci\'on
Banco Exterior, Spain.}  \\
\vskip .3cm
{\large M. Quir\'os}$^{1,}$\footnote{Work partly supported by CICYT, Spain,
under contract AEN90-0139.}
\\
\vskip .3cm
and \\
\vskip .3cm
{\large F. Zwirner}\footnote{On leave from INFN, Sezione di Padova, Padua,
Italy.}
\\
\vskip .3cm
Theory Division, CERN, \\
Geneva, Switzerland \\
\vskip 1cm
\end{center}
\begin{abstract}
\noindent
The basic tool for the study of the electroweak phase transition is
$\veff$, the one-loop finite-temperature effective potential, improved
by all-loop resummations of the most important infrared contributions. In
this paper we perform, as a first step towards a full analysis of the
Standard Model case, a detailed study of the effective potential of the
scalar theory. We show that subleading corrections to the self-energies
lead to spurious terms, linear in the field-dependent mass $m(\phi)$,
in the daisy-improved effective potential.
Consistency at subleading order requires the introduction of superdaisy
diagrams, which prevent the appearance of linear terms. The resulting
$\veff$ for the scalar theory hints at a phase transition which is either
second-order or very weakly first-order.
\end{abstract}
\vfill{
CERN-TH.6451/92
\newline
\noindent
April 1992}
\end{titlepage}
\setcounter{footnote}{0}
\vskip2truecm

It was argued long ago by Kirzhnits and Linde [\ref{kirlin}] that the
electroweak gauge symmetry should be restored at sufficiently high
temperature. Soon after this, Weinberg [\ref{weinberg}] and Dolan and Jackiw
[\ref{dj}] laid the foundations for a quantitative treatment of this problem
(for reviews of later developments, see e.g. refs.~[\ref{books}]).
Recently, the observation [\ref{sphaleron}] that the rate of anomalous
$B$-violating processes is unsuppressed at high temperatures has revived
the interest in the subject: a detailed discussion of the electroweak phase
transition is required, if one wants to confront any model of particle
interactions with the observed cosmological baryon asymmetry.
In the Standard Model (SM), the amount of CP violation appears to be too
small for baryogenesis at the electroweak scale (for recent reviews,
see e.g. refs.~[\ref{reviews}]). However, the requirement that a pre-existing
baryon asymmetry be not washed out by anomalous $B$-violating interactions,
just after the electroweak phase transition, can in principle put a stringent
upper bound on the SM Higgs mass (tentative bounds have been given in
refs.~[\ref{protoshapo}], and possible ways out discussed in
refs.~[\ref{waysout}]). In extensions of the SM, one can aim at the
construction of consistent and phenomenologically acceptable models for
baryogenesis at the electroweak scale, even if a detailed description of
the latter has to face a number of technical difficulties [\ref{reviews}].

The basic tool for the discussion of the electroweak phase transition is
$V_{eff}(\phi,T)$, the finite-temperature one-loop effective potential,
improved by all-loop resummations of the most important infrared contributions.
A lively theoretical debate [\ref{hall}--\ref{arnold}] on the structure of
$\veff$ has recently taken place, with a certain amount of disagreement. In
particular, the existence and the nature of terms linear in the field-dependent
masses, in the high-temperature expansion of $\veff$ and after the inclusion
of subleading infrared corrections, are rather controversial: a positive
linear term was found in [\ref{shapo}], whilst a negative linear term was
obtained in [\ref{brahm}]. It was subsequently argued in [\ref{dine}], on
the basis of general arguments, that there are no linear terms, but no
explicit proof of this statement at subleading order was given.
To better understand the origin of these conflicting results and the nature
of the electroweak phase transition in the SM, we discuss here some of the
issues in a simpler context, the theory of a single real scalar. The
present analysis will be extended to the more complicated case of the
Standard Model in a longer paper, currently in preparation. After reviewing
some well-known results, we explicitly compute the combinatorics of tadpole
and vacuum daisy diagrams, which was recently questioned in [\ref{dine}].
We show that by integrating the tadpoles one precisely finds the one-loop
effective potential, with the field-dependent mass $m^2(\phi)$ replaced,
in the infrared-dominated terms, by an effective $T$-dependent mass
$\mbar^2 (\phi,T)$, in agreement with the result obtained by direct
calculation of the vacuum daisy diagrams. We also improve over the previous
computations by showing how to include subleading terms in the computation
of the effective mass $\mbar^2 (\phi,T)$. We show that, to include them
consistently, avoiding at the same time the problem of negative squared
masses, one must go to the superdaisy approximation, where $\mbar^2(\phi,T)$
is computed as the self-consistent solution of a corresponding gap equation.
We finally solve the gap equation and study the behaviour of the resulting
effective potential: within the limits of our improved perturbative
expansion, we find that the phase transition is either second-order or
very weakly first-order.

We consider the theory of a real scalar field $\phi$, with a tree-level
potential (invariant under the discrete ${\bf Z}_2$ symmetry $\phi \to
- \phi$)
\be
\label{vtreephi}
V_{tree}(\phi) = - {\mu^2 \over 2} \phi^2 + {\lambda \over 4} \phi^4,
\ee
and positive $\lambda$ and $\mu^2$. At the tree level, the
field-dependent mass of the scalar field is $m^2(\phi) = 3
\lambda \phi^2 - \mu^2$, and the minimum of $V_{tree}$
corresponds to $\phi^2 = \mu^2 / \lambda \equiv v^2$, so that $m^2(v) = 2
\lambda v^2 = 2 \mu^2$. The basic tool for the study of the vacuum state of
the theory is the effective potential [\ref{effpot}]. To perform a systematic
loop expansion of the effective potential at finite temperature, it
is convenient, following ref.~[\ref{dj}], to shift the scalar field
around a constant background value $\overline{\phi}$~: $\phi =
\overline{\phi} + \phi'$. Since the procedure is well known, and there
is no risk of confusion, in the following we shall use the symbol
$\phi$ also for $\overline{\phi}$. All our calculations will be performed
in the imaginary-time formalism.

The one-loop contribution to the effective potential, $V_{eff} = V_{tree}
+ V^{(1)} + V^{(2)} + \ldots$~, comes from the quadratic terms in $\phi '$
in the effective action, and is given, at the temperature $T$, by [\ref{dj}]
\be
\label{v1}
V^{(1)} [ m^2 (\phi) , T ] =
{T \over 2} \sum_n \int
{d^3 \vec{p} \over {(2 \pi)^3}}
\log \left[  \omega_n^2 + \vec{p}^{\; 2} + m^2(\phi) \right] \, ,
\ee
where $\omega_n = 2 \pi n T$ ($n \in {\bf Z}$) are the bosonic Matsubara
frequencies, and field-independent contributions will be neglected here and
in the following. It was stressed in refs.~[\ref{dine}] that $V^{(1)}$ can
be easily obtained as
\be
\label{alternative}
V^{(1)} =
\int d \phi \,
{\cal T}^{(1)},
\ee
where ${\cal T}^{(1)}$ is the one-loop tadpole diagram. Indeed, using
the Feynman rules associated to (\ref{vtreephi}) and including the
appropriate symmetry factor, we can write
\be
\label{vtad1}
{\cal T}^{(1)} [ m^2 (\phi) , T ] =
6 \lambda \phi \, {T \over 2} \sum_n
\int
{d^3 \vec{p} \over {(2 \pi)^3}}
{1 \over {\omega_n^2 + \vec{p}^{\; 2} + m^2(\phi)}} \, ,
\ee
which, upon integration over $\phi$, gives eq.~(\ref{v1}).

Before moving to the discussion of higher-loop contributions to the
effective potential, we recall some well-known results, which will be
useful later on. Following ref.~[\ref{dj}], we can split the one-loop
contribution of eq.~(\ref{v1}) into a $T=0$ and a $T \ne 0$ part
\be
\label{djsplit}
V^{(1)} [ m^2 (\phi) , T ] =
V^{(1)} [ m^2 (\phi) , 0 ] +
\Delta V^{(1)} [ m^2 (\phi) , T ] \, .
\ee
After introducing an appropriate set of counterterms, the
$T=0$ contribution reads
\be
\label{v1zerobis}
V^{(1)} [ m^2 (\phi) , 0 ] =
{m^4(\phi) \over 64 \pi^2} \left[  \log {m^2(\phi) \over 2 \mu^2}
- {3 \over 2} \right]  \, .
\ee
The specific form of the counterterms, and therefore of $V^{(1)} [ m^2
(\phi) , 0 ]$, depends on the renormalization prescription.
Eq.~(\ref{v1zerobis}) corresponds to the $\overline{MS}$ scheme
[\ref{msbar}], with a renormalization scale $Q^2=2 \mu^2$.
The $T \ne 0$ contribution is given by
\be
\label{deltav1t}
\Delta V^{(1)} [ m^2 (\phi) , T ] = {T^4 \over 2 \pi^2} J_+ (y^2) \, ,
\ee
where
\be
\label{ypsilon}
y^2 \equiv { m^2 (\phi) \over T^2 } \, ,
\;\;\;\;\;
J_+ (y^2) \equiv \int_0^{\infty} dx \, x^2 \,
\log \left( 1 - e^{- \sqrt{x^2 + y^2}} \right) \, .
\ee
In the high-temperature limit $y^2 \ll 1$, we can write the expansion
\be
\label{dv1tht}
\Delta V^{(1)} [ m^2 (\phi) , T ] =
{\displaystyle{m^2 (\phi) T^2 \over 24}}
- {\displaystyle{m^3 (\phi) T \over 12 \pi}}
{\displaystyle
- {m^4(\phi) \over 64 \pi^2}  \left[ \log {m^2(\phi) \over T^2} - 5.4076
\right] + \ldots } \, ,
\ee
where the dots stand for terms $O[m^6(\phi)/T^2]$ or higher.
Observe the cancellation of the $m^4 \log m^2$ terms between the $T=0$
and the $T \ne 0$ contributions in the high-temperature expansion.

The simplest approximation to the $T$-dependent effective
potential, $\veff$, consists in adding to $V_{tree}$ only the
leading (field-dependent) contribution to $\v1t$ in the high-temperature
expansion. As a result, we obtain
\be
\label{brute}
{\displaystyle
\veff = \left( - \mu^2 + {\lambda T^2 \over 4} \right) {\phi^2
\over 2} + {\lambda \over 4} \phi^4 } \, ,
\ee
whose behaviour is illustrated in fig.~1a, for the representative choice of
parameters $\lambda = 0.1$, $v = 250$ GeV. One can easily see that
eq.~(\ref{brute}) describes a second-order phase transition, with the
critical temperature given by $T_0 = 2 \mu / \sqrt{\lambda} = 2 v$,
and the $T$-dependent minimum $v_T$ evolving as $v_T^2 = v^2 - T^2/4$
for $T < T_0$.

If, instead of the approximation (\ref{brute}), one wants to use the full
one-loop effective potential, $\veff = V_{tree}(\phi) + \v1t$,
one runs into several difficulties. To begin with, for $\phi < v /
\sqrt{3}$ the field-dependent mass $m^2 (\phi)$ is negative, and some
expressions in eqs.~(\ref{v1})--(\ref{dv1tht}) become ill-defined.
For $T$ small enough, the minimum $v_T$ of the one-loop effective
potential is still large enough that in its neighbourhood $m^2(\phi)>0$.
However, for temperatures close to $T_0$ this cannot be
true, and the na{\"\i}ve expression of the one-loop effective potential
is of no use. The other well-known problem of the na{\"\i}ve one-loop
potential is the breakdown of the perturbative expansion near the
critical temperature, which is due to the presence of higher-loop
infrared-divergent diagrams in the limit $m^2(\phi) \to 0$. Both of these
problems can be alleviated by the use of resummation techniques,
which take into account the most important infrared contributions to
the effective potential at all orders in the perturbative expansion.
These resummations correct the part of the one-loop
effective potential which is associated to the modes of zero Matsubara
frequency, $\omega_n = 0$: a quick glance at eqs.~(\ref{alternative})
and (\ref{vtad1}) shows that it corresponds precisely to the $m^3$ term
in the high-temperature expansion of eq.~(\ref{dv1tht}).

We now recall how to perform the resummation of the most important class of
infrared-dominated diagrams, the so-called daisy and superdaisy diagrams.
In contrast to ref.~[\ref{dj}] (and to most of the subsequent refinements
[\ref{fendley},\ref{origin}]), we shall work with just one scalar field $\phi$
and for an arbitrary constant background field configuration, not only for
$\phi=0$. We shall justify {\em a posteriori} our procedure, after having
properly identified the parameters of the improved perturbative expansion.

As a first step, we need to compute the one-loop scalar self-energy
at finite temperature, corresponding to the two diagrams depicted in
fig.~2, in the infrared limit $p^0=0$, $\vec{p} \to 0$, where $p = (p^0,
\vec{p} \, )$ is the external momentum.
For the temperature-dependent self-energy we find
\be
\label{piscalar}
\Pi [ m^2 (\phi) , T ] =
\Pi^{(a)} [ m^2 (\phi) , T ]
+ \Pi^{(b)} [ m^2 (\phi) , T ] \, ,
\ee
where, after ($\ov{MS}$) subtraction of the appropriate $T=0$
counterterms\footnote{
Notice that to obtain $\Pi [m^2(\phi),T]$ it is not necessary to perform any
new diagrammatic calculation, since to compute the self-energy at zero
external momentum one can simply take the second derivative
of $V^{(1)} [m^2(\phi),T]$ in eq.~(\ref{djsplit}).},
\be
\label{pi1}
\Pi^{(a)} [ m^2 (\phi) , T ] =
{1 \over 32 \pi^2} \left[
6 \lambda m^2 (\phi) \left( \log {m^2(\phi) \over 2 \mu^2} - 1 \right)
\right] + {3 \lambda T^2 \over 2 \pi^2} \cdot I_+ ( y^2 ) \, ,
\ee
\be
\label{pi2}
\Pi^{(b)} [ m^2 (\phi) , T ] =
{1 \over 32 \pi^2} \left[
36 \lambda^2 \phi^2 \log {m^2(\phi) \over 2 \mu^2}
\right] + {9 \lambda^2 \phi^2 \over \pi^2} \cdot I_+' ( y^2 ) \, ,
\ee
and
$I_+ ( y^2 ) = 2 [d J_+ (y^2)/ d y^2] \,$,
$I_+' ( y^2 ) = [d I_+ (y^2) / d y^2] \,$.

In the calculation of the effective potential, the second step of the daisy
resummation procedure consists in replacing, in the one-loop terms originating
from the zero-frequency modes, i.e. in the $m^3$ term of the high-temperature
expansion (\ref{dv1tht}), the field-dependent mass $m^2(\phi)$ by an
effective, $\phi$- and $T$-dependent mass,
\be
\label{mbardai}
\mbar^2 (\phi,T) = m^2 (\phi) + \Pi [m^2(\phi),T] \, ,
\ee
where $\Pi [m^2(\phi),T]$ is the one-loop self-energy in the infrared limit,
given by eqs.~(\ref{piscalar})--(\ref{pi2}). The standard way of
implementing this
substitution has been recently questioned in ref.~[\ref{dine}], so we give
here some details of the calculation. Both techniques previously discussed
at the one-loop level can be applied to obtain the effective potential at
any order in the perturbative expansion, and in particular to the resummation
of daisy diagrams.

The daisy diagrams contributing to the tadpole ${\cal T}_0$
(where the subscript zero means that only zero-frequency modes are propagating
in big bubbles) are depicted in figs.~3a--d. Considering all inequivalent
permutations of bubbles, we can write
\be
\label{taddai}
{\cal T}^{daisy}
\equiv \sum_{i=1}^{4} {\cal T}^{(d_i)}
= \sum_{i=1}^{4} \sum_{j,l} C^{(d_i)}_{j,l}
T^{(d_i)}_{j,l} \, ,
\ee
where the combinatorial factors are given by
$$
C^{(d_1)}_{j,l} =
\left( \begin{array}{c} j + l \\ l \end{array} \right)
2^{-(j+l+1)} \, ,
\;\;\;\;\;
C^{(d_2)}_{j,l}  =
\left( \begin{array}{c} j + l - 1 \\ l \end{array} \right)
2^{-(j+l+1)} \, ,
$$
\be
\label{combi}
C^{(d_3)}_{j,l} = C^{(d_4)}_{j,l} =
\left( \begin{array}{c} j + l - 1 \\ j \end{array} \right)
2^{-(j+l)}.
\ee
Using the Feynman rules associated to (\ref{vtreephi}) and the self-energies
of fig.~2, we can write
\be
\label{v1tad}
{\cal T}^{(d_1)} \equiv
\sum_{j,l} C^{(d_1)}_{j,l} T^{(d_1)}_{j,l}
= {\cal T}_0^{(1)} (m^2 + \Pi) - {\cal T}_0^{(1)} (m^2),
\ee
where ${\cal T}_0^{(1)}$ is the zero-frequency part of the one-loop tadpole
defined in eq.~(\ref{vtad1}). We then see that adding the class of diagrams
in fig.~3a to the one-loop tadpole diagram amounts to shifting $m^2$ by $\Pi$
in the zero-frequency part of the propagator. Eq.~(\ref{v1tad})
cannot be integrated to give a logarithm, because $\Pi$ depends in general
\footnote{Only by making an approximation where $\Pi$ does not depend on
$\phi$, as in refs.~[\ref{carrington},\ref{dine}], eq.~(\ref{v1tad}) can be
integrated in $\phi$, giving as a result a logarithmic function.} on $\phi$.
However, the previous result changes when one considers the additional
contributions to ${\cal T}^{daisy}$ coming from figs.~3b--d. It is easy
to check that
\be
\label{v2tad}
{\cal T}^{(d_2)} =
{d \Pi^{(a)} \over {d \phi}}
{T \over 2} \int
{d^3 \vec{p} \over {(2 \pi)^3}}
{1 \over {\vec{p}^{\; 2} + m^2(\phi) + \Pi}} \, ,
\ee
\be
\label{v34tad}
{\cal T}^{(d_3)} + {\cal T}^{(d_4)} =
{d \Pi^{(b)} \over {d \phi}}
{T \over 2} \int
{d^3 \vec{p} \over {(2 \pi)^3}}
{1 \over {\vec{p}^{\; 2} + m^2(\phi) + \Pi}} \, .
\ee
Adding up the one-loop and daisy contributions to the tadpole,
eqs.~(\ref{vtad1}) and  (\ref{v1tad})--(\ref{v34tad}), we obtain
\be
\label{vdaitad}
{\cal T}_0^{(1)} + {\cal T}^{daisy} =
\left( 6 \lambda \phi + {d \Pi \over {d \phi}}
\right) {T \over 2} \int
{d^3 \vec{p} \over {(2 \pi)^3}}
{1 \over {\vec{p}^{\; 2} + m^2(\phi) + \Pi}} \, ,
\ee
which yields for the effective potential, upon integration over $\phi$,
\be
\label{veffdaisy}
\veff = V_{tree}(\phi) + \ov{V}^{(1)} \, ,
\ee
\be
\label{delirium}
\ov{V}^{(1)}
\equiv
V^{(1)} [m^2 (\phi),T]
-{T \over 12 \pi} [\mbar^3 (\phi,T) - m^3 (\phi)] \, ,
\ee
where $V^{(1)}$ was given in eq.~(\ref{v1}), $\mbar^2$ in eq.~(\ref{mbardai}),
and $\ov{V}^{(1)}$ is the daisy-improved one-loop contribution to the
effective potential. We would like to stress again two important points:
1) only the $m^3$ term, which arises from the infrared-dominated diagrams,
receives the shift $m^2 \to \mbar^2$; 2) the diagrams in figs.~3b--d are
essential to obtain the result of eq.~(\ref{veffdaisy}), unless one is
working in an approximation where $\Pi$ does not depend on $\phi$, in
which case they vanish\footnote{We believe that
this is the origin of the apparent discrepancy found in ref.~[\ref{dine}].
This discrepancy disappears when all diagrams are included.}.

The result of eq.~(\ref{veffdaisy}) can be obtained directly
from the diagrammatic expansion of the vacuum daisy diagrams shown in
fig.~3e, where the small bubbles are taken at zero external momentum, and
only zero-frequency modes are propagating in the big bubbles.
Adding all inequivalent permutations of bubbles,
\be
\label{vdaisy}
V^{daisy} = \sum_{j,l} C_{j,l} \, V_{j,l} \, ,
\ee
where the combinatorial factors are
\be
\label{combfac}
C_{j,l} = {(j + l - 1) ! \over {j ! \, l !}} \,  2^{- (j + l + 1)},
\ee
so that
\be
\label{vdaibis}
V^{daisy} = -{T \over 2} \sum_{N=1}^{\infty} {1 \over N} \int
{d^3 \vec{p} \over {(2 \pi)^3}}
{(- \Pi )^N \over [\vec{p}^{\; 2} + m^2(\phi)]^N }
\, .
\ee
More explicitly, $V_{eff} = V_{tree}+  V^{(1)} (m^2) + V^{daisy} = V_{tree}
+ \ov{V}^{(1)}$, which is precisely eq.~(\ref{veffdaisy}).

The combinatorics of two-loop graphs, both for vacuum and for tadpole diagrams,
do not fit the general rules given in (\ref{combfac}) and (\ref{combi}),
respectively. However, they do so if only the zero frequency mode propagates
in the main loop, as in the case under consideration. In particular, the first
two-loop diagram in the table comes with a combinatorial factor of $1/8$,
whilst eq.~(\ref{combfac}) gives a factor of $1/4$. The extra factor of 2
comes from the symmetry of the diagram, since both bubbles are equivalent
and the propagation of the zero-frequency mode can be considered for each
of the two. The second two-loop diagram in the table comes with a
combinatorial factor of $1/12$, whilst again $1/4$ is required by
eq.~(\ref{combfac}). The extra factor of 3 comes again from the
symmetry of the diagram, since the three propagators are equivalent and
each of them can carry the zero-frequency mode.

We can now study the effective potential at the level of the
daisy approximation, eq.~(\ref{veffdaisy}). In
particular, we can study the effects of the $m^3$ term in
the high-temperature expansion of eq.~(\ref{dv1tht}), which hints
at the possibility of a first-order phase transition.
The simplest improvement over the na{\"\i}ve one-loop effective potential
consists in keeping only the leading term in the high-temperature expansion
of the self-energy, i.e. to take $\Pi (m^2) = (\lambda/4)
T^2$, as was done in ref.~[\ref{carrington}]. In that case the effective
mass to be substituted in eq.~(\ref{veffdaisy}) reads
\be
\label{carsol}
\mbar^2 (\phi, T) = m^2(\phi) + {\lambda \over 4} T^2 \, ,
\ee
which becomes $3 \lambda \phi^2$ at $T_0 = 2 v$. At that temperature there is
a $\phi^3$ term (with negative coefficient) in the high-temperature expansion
of (\ref{veffdaisy}), while the coefficient of the $\phi^2$ term vanishes.
This is the signal of the onset of a first-order phase transition: for $T$
just above $T_0$, the potential has a positive curvature at $\phi=0$, but the
effect of the $\phi^3$ term bends down the potential at moderate values of
$\phi$. We show the effective potential corresponding to this approximation
in fig.~1b, and the values of the effective mass (\ref{carsol}) in fig.~4a.

To further improve over the approximation (\ref{carsol}), we could
try to perform in eq.~(\ref{veffdaisy}) the full shift defined by
eq.~(\ref{mbardai}). However, in the
calculation of $\Pi[m^2(\phi),T]$ we still have to face the problem of
negative squared masses, which does not allow us to explore $\phi^2 < v^2 / 3$.
Disregarding this problem for a moment, we can look at the
behaviour of the $m^3$ term in the high-temperature expansion (\ref{dv1tht})
of $V^{(1)}$. Performing the shift $m^2 \to \mbar^2$, with $\mbar^2$ given
by (\ref{mbardai}), the $m^3$ term in $V^{(1)}$ would become
\be
\label{rise}
- {T \over {12 \pi}}
\left[ m^2 + \Pi (m^2) \right]^{3/2} \, ,
\ee
which would give rise, in the high-temperature expansion
\be
\label{hight}
\Pi [ m^2 (\phi) , T ] = {\lambda \over 4} T^2 - {3 \lambda \over
{4 \pi}} m T - {9 \lambda^2 \over {4 \pi}} \phi^2 {T \over m} + \ldots
\, ,
\ee
to a term linear in $m$. This linear term originates from {\em subleading}
contributions to the self-energy: similar terms have recently been obtained
in the SM effective potential [\ref{shapo},\ref{brahm}], and will be discussed
in that context in our forthcoming paper. Both in the scalar theory and in
the SM, the linear term is an artefact of the resummation of daisy
diagrams\footnote{Independent arguments for the absence of linear terms,
based on effective-theory methods, have been given in [\ref{dine}].}.
As discussed in more detail later, when including subleading terms in the
infrared improvement of the effective potential, we should also include
the so-called superdaisy diagrams [\ref{dj}]. Including the
superdaisy diagrams amounts to solving the gap equation
\be
\label{gap}
\mbar^2 (\phi,T) = m^2 (\phi) + \Pi [\mbar^2(\phi,T),T] \, ,
\ee
and to substitute its solution in eq.~(\ref{veffdaisy}).

Before proceeding, a few words about the expansion parameters corresponding
to the different approximations are in order. The leading behaviour of two-
and three-loop corrections to the zero-frequency part of the one-loop effective
potential is presented in the table. The reader can easily extend the
classification to higher loop orders. The relevant parameters are
[\ref{fendley}]
\be
\label{alphat}
\alpha \equiv \lambda {T^2 \over m^2} \, ,
\;\;\;\;\;
\beta \equiv \lambda {T \over m} \, .
\ee
Subleading corrections have an extra
suppression factor $(\beta/\alpha)$ with respect to the leading behaviour.
Since we are working at arbitrary values of $\phi$, we must also introduce
\be
\label{gammat}
\gamma \equiv {\phi^2 \over T^2} \, ,
\ee
which comes from the trilinear vertices.
The validity of the (unimproved) loop expansion requires not only
$\lambda < 1$, but also the additional condition $\alpha < 1$.
For values of $\phi$ and $T$ that are relevant for the description of the
phase transition, $\alpha \simgt 1$ and the perturbative expansion in
$\alpha$ breaks down [\ref{weinberg},\ref{dj},\ref{fendley}], but even in
this case $\beta$ can still be a small expansion parameter.
The daisy resummation amounts to replacing $m^2$ by $\mbar^2$, given by
eq.~(\ref{mbardai}), in the perturbative expansion, in which case
$\bar{\alpha} \equiv \lambda T^2 / \mbar^2 \simgt 1$ in the region of interest
for the phase transition, while $\bar{\beta}$ may remain as
a good expansion parameter. As for the parameter $\gamma$ in
eq.~(\ref{gammat}),
if we consider values of $\phi$ such that $\phi \ll T$, then $\gamma \ll 1$
and diagrams proportional to $\gamma^n$ ($n \ge 1$) can be
neglected\footnote{This is the case of diagrams involving trilinear
couplings. These diagrams were not considered in previous calculations,
performed at $\phi=0$.}. Here we first assume that $\bar{\alpha} \bar{\beta}
\gamma \ll 1$ (as could be the case in the region of interest for the study
of the phase transition), so that they can be consistently neglected in the
resummation procedure. The latter prescription amounts to dropping
the $\Pi^{(b)}$ term in the second member of the gap equation (\ref{gap}).
{}From the table we can see that diagrams not belonging to the daisy or
superdaisy classes
are suppressed by factors of order at least $\bar{\beta}^2$, so that
we can keep consistently in the gap equation terms of order $\bar{\beta}$,
i.e. subleading terms from $\Pi^{(a)}$. Keeping in general the subleading
terms in the self-energy (\ref{hight}) amounts to working to $O(\bar{\beta})$,
as we are doing in this paper, whereas the approximation of
eq.~(\ref{carsol}) amounts to working to
$O(\bar{\beta}^0)$. Working to $O(\bar{\beta})$ is necessary to include
linear terms in the self-energy, and it is the best one can afford
without considering the non-daisy, non-superdaisy diagrams (see table).
The validity of both approximations relies upon the condition $\bar{\beta}
< 1$.

To obtain an approximate analytical solution to the gap equation (\ref{gap}),
we can write its high-temperature expansion (it is not restrictive to assume
$\mbar \ge 0$)
\be
\label{pihight}
\mbar_0^2 (\phi, T) = m^2 (\phi) + {\lambda \over 4} T^2 -
{3 \lambda T \over {4 \pi}} \mbar_0 (\phi, T) \, ,
\ee
where the zero subscript reminds us that we are neglecting the $\phi^2$ term
in (\ref{hight}). Eq.~(\ref{pihight}) admits the obvious solution, valid up
to $O(\bar{\beta}^2)$ corrections,
\be
\label{solht}
\mbar_0(\phi,T) =
- {3 \lambda T \over 8 \pi}
+ \sqrt{
\left( {3 \lambda T \over 8 \pi} \right)^2
+ \lambda {T^2 \over 4} + m^2 (\phi) }
\, .
\ee
At $T \ge T_0$, $\mbar_0(0,T) \ge 0$, while at $T < T_0$ there is no solution
for values of $\phi$ such that $\phi^2 < (v^2 - T^2/4)/3$. From the explicit
solution (\ref{solht}) we can see that $\mbar_0$ is an analytic function of
$m^2$ near $m^2=0$, in contrast with the daisy approximation (\ref{mbardai}),
so that there can be no terms linear in $m(\phi)$ in the effective potential.
Therefore, as anticipated, the absence of linear terms from the improved
one-loop effective potential (at subleading order) is ensured by the
consistent inclusion of the superdaisy infrared-dominated diagrams.

The exact solution to the gap equation (\ref{gap}), with $\Pi = \Pi^{(a)}$,
is plotted in fig.~4b
[the approximation (\ref{solht}) gives results which are indistinguishable on
the scale of fig.~4b], for different temperatures close to the critical one.
We also display in fig.~4c the corresponding values of the parameters
$\bar{\alpha}/(2 \pi)$ (dashed lines) and $\bar{\beta}/(2 \pi)$ (solid lines).
They give indications on the values of $\phi$ and $T$
at which our (superdaisy-improved) perturbative expansion breaks down.
The resulting effective potential (\ref{veffdaisy}) is displayed in fig.~1c.

A word of caution should be said here, since the validity of our last
result relies upon that of the $\bar{\beta}$ expansion.
As shown in fig.~4c, $\bar{\beta}/(2 \pi)$ is of order one for the values
of $T$ and $\phi$ associated to the apparent first-order structure.
Furthermore, we have also plotted in fig.~4c the parameter $\bar{\alpha}
\bar{\beta} \gamma$ (dash-dotted lines), which measures how much the
diagrams involving
trilinear couplings, and so the $\Pi^{(b)}$ term in eq.~(\ref{gap}),
are negligible. We can see from fig.~4c that the condition $\bar{\alpha}
\bar{\beta} \gamma \ll 1$ is not fulfilled in the relevant region, this
not being unrelated to the fact that also the condition $\bar{\beta}
\ll 1$ is not fulfilled in that region.

Although we know, according to our previous arguments, that consistency
of the superdaisy approximation, which is an expansion in the parameter
$\bar{\beta}$ neglecting $O(\bar{\beta}^2)$ corrections, requires $\Pi^{(b)}$
not to be relevant in the gap equation (\ref{gap}), we now include
it for the sake of comparison. We are then led to solving the gap equation
\be
\label{gapphi}
\mbar^2 (\phi, T) = m^2 (\phi) + {\lambda \over 4} T^2
- {3 \lambda T \over {4 \pi}} \mbar (\phi, T)
- {9 \lambda^2 \over {4 \pi}} {\phi^2 T \over  \mbar (\phi, T)}
\, ,
\ee
where a high-temperature expansion has already been performed.
Since our expansions hold up to corrections  $O(\bar{\beta}^2)$,
we can consistently  give the analytic solution of eq.~(\ref{gapphi}),
to order $\bar{\beta}$, as
\be
\label{solphi}
\mbar^2(\phi,T) =
\mbar_0^2(\phi,T)
- {9 \lambda^2 \over {4 \pi}} {\phi^2 T \over
\sqrt{ m^2 (  \phi ) + {\lambda T^2 \over 4}}}
\, ,
\ee
where $\mbar_0(\phi,T)$ is the solution (\ref{solht}). The solution
(\ref{solphi}) to the gap equation is plotted in fig.~4d, and the
resulting effective potential is displayed in fig.~1d. From
eq.~(\ref{solphi}) we can see, as was the case from eq.~(\ref{solht}),
that the absence of terms linear in $m(\phi)$ from the effective
potential follows from the resummation of the superdaisy diagrams.
{}From fig.~4d we can see that the gap equation (\ref{gapphi})
does not have a real solution for a range of $\phi$ values
which depends on the temperature $T$. Inside this range the solution
(\ref{solphi}) becomes purely imaginary, and so the $\mbar^3$ term of the
effective potential becomes purely imaginary as well. This situation
was discussed in general in [\ref{wwu}], where it was shown that the
imaginary part of the effective potential has a natural interpretation
as a decay rate per unit volume of a well-defined quantum state. We shall
disregard the contribution from those imaginary-frequency modes and plot the
real part of the effective potential in fig.~1d. Solid (dash-dotted) lines
correspond to the region where the gap equation does (does not) have a real
solution and the effective potential does not (does) get an imaginary part.
Fig.~1d hints, as was the case in the
approximation of fig.~1c, at a very weakly first-order phase transition.
However, this structure appears again in the region where even our
improved perturbative expansion is not reliable, as one can easily read
from fig.~4c. Therefore we can only
conclude that the phase transition is either second-order or
extremely weakly first-order.
Our findings are in agreement with general results based on the
$\epsilon$-expansion [\ref{ginsparg}], which suggest second-order
phase transitions in theories without an asymptotically-free coupling.

In conclusion, we have analysed the structure of the one-loop effective
potential for the scalar theory, when higher-order infrared-dominated diagrams
are resummed to leading and subleading order. We have explicitly shown that
the combinatorics of higher-loop diagrams is appropriate to shift the
value of $m^2(\phi)$ in the zero-frequency part of the one-loop effective
potential. We have identified the parameters of the improved expansion, and
proved that including subleading terms in the self-energy is {\em not}
consistent with keeping only the daisy class of diagrams: such a procedure
would produce spurious linear terms in $m(\phi)$ in the effective potential.
Consistency at subleading order requires introducing superdaisy diagrams,
which prevent the appearance of linear terms. In the scalar theory, the
improved effective potential at subleading order suggests, within the
limits of the improved perturbative expansion, a phase transition
which is either second-order or very weakly first-order.

\section*{Acknowledgements}
We acknowledge discussions with T.~Altherr, A.~Brignole, J.~Orloff, A.~Rebhan
and M.E.~Shaposhnikov.

\newpage
\section*{References}
\begin{enumerate}
\item
\label{kirlin}
D.A.~Kirzhnits and A.D.~Linde, Phys. Lett. 72B (1972) 471.
\item
\label{weinberg}
S. Weinberg, Phys. Rev. D9 (1974) 3357.
\item
\label{dj}
L. Dolan and R. Jackiw, Phys. Rev. D9 (1974) 3320.
\item
\label{books}
D.J.~Gross, R.D.~Pisarski and L.G.~Yaffe,  Rev. Mod. Phys. 53 (1981) 43;
\\
J.I.~Kapusta, {\em Finite-temperature Field Theory} (Cambridge University
Press, 1989);
\\
A.D. Linde, {\em Particle Physics and Inflationary Cosmology} (Harwood,
New York, 1990).
\item
\label{sphaleron}
F.R.~Klinkhammer and N.S.~Manton, Phys. Rev. D30 (1984) 2212;
\\
V.A.~Kuzmin, V.A.~Rubakov and M.E.~Shaposhnikov, Phys. Lett.
B155 (1985) 36;
\\
P.~Arnold and L.~McLerran, Phys. Rev. D36 (1987) 581 and D37 (1988) 1020;
\\
M.~Dine, O.~Lechtenfeld, B.~Sakita, W.~Fischler and J.~Polchinski,
Nucl. Phys. B342 (1990) 381.
\item
\label{reviews}
A.D.~Dolgov, Kyoto preprint YITP/K-940 (1991);
\\
M.E.~Shaposhnikov, preprint CERN-TH.6304/91;
\\
M.~Dine, Santa Cruz preprint SCIPP 91/27.
\item
\label{protoshapo}
M.E.~Shaposhnikov, JETP Lett. 44 (1986) 465, Nucl. Phys. B287 (1987) 757
and B299 (1988) 797;
\\
A.I. Bochkarev, S.Yu. Khlebnikov and M.E.~Shaposhnikov, Nucl. Phys. B329
(1990) 490;
\\
A.I. Bochkarev, S. Kuzmin and M.E.~Shaposhnikov, Phys. Lett. B244 (1990) 27.
\item
\label{waysout}
M.~Fukugita and T.~Yanagida, Phys. Rev. D42 (1990) 1285;
\\
J.A.~Harvey and M.S.~Turner, Phys. Rev. D42 (1990) 3344;
\\
A.E.~Nelson and S.M.~Barr, Phys. Lett. B246 (1991) 141;
\\
B.A.~Campbell, S.~Davidson, J.~Ellis and K.A.~Olive, Phys. Lett B256 (1991)
457;
\\
W.~Fishler, G.F.~Giudice, R.G.~Leigh and S.~Paban, Phys. Lett. B258 (1991) 45;
\\
L.E.~Ib\'a\~nez and F.~Quevedo, preprint CERN-TH.6433/92;
\\
H.~Dreiner and G.G.~Ross, Oxford preprint OUTP-92-08P.
\item
\label{hall}
G.W.~Anderson and L.J.~Hall, Phys. Rev. D45 (1992) 2685.
\item
\label{shapo}
M.E.~Shaposhnikov, Phys. Lett. B277 (1992) 324.
\item
\label{brahm}
D.E.~Brahm and S.D.H.~Hsu, Caltech preprints CALT-68-1762,
HUTP-91-A064 and CALT-68-1705, HUTP-91-A063.
\item
\label{carrington}
M.E. Carrington, Phys. Rev. D45 (1992) 2933.
\item
\label{dine}
M.~Dine, R.G.~Leigh, P.~Huet, A.~Linde and D.~Linde,
Santa Cruz preprint SCIPP-92-06, SLAC-PUB-5740, SU-ITP-92-6 and
SLAC preprint SLAC-PUB-5741, SCIPP-92-07, SU-ITP-92-7.
\item
\label{arnold}
P.~Arnold, Univ. of Washington preprint UW/PT-92-06, NUHEP-TH-92-06.
\item
\label{effpot}
J. Iliopoulos, C. Itzykson and A. Martin, Rev. Mod. Phys. 47 (1975) 165,
and references therein;
\\
S. Coleman and E. Weinberg, Phys. Rev. D7 (1973) 1888;
\\
S. Weinberg, Phys. Rev. D7 (1973) 2887;
\\
R. Jackiw, Phys. Rev. D9 (1974) 1686.
\item
\label{msbar}
W.A.~Bardeen, A.J.~Buras, D.W.~Duke and T.~Muta, Phys. Rev. D18 (1978) 3998.
\item
\label{fendley}
P.~Fendley, Phys. Lett. B196 (1987) 175.
\item
\label{origin}
K.~Takahashi, Z. Phys. C26 (1985) 601;
\\
T.~Altherr, Phys. Lett. B238 (1990) 360;
\\
N.~Banerjee and S.~Mallik, Phys. Rev. D43 (1991) 3368;
\\
R.R.~Parwani, Stony Brook preprint ITP-SB-91-64;
\\
V. Jain, Max-Planck-Institute preprint MPI-Ph/92-41.
\item
\label{wwu}
E.J.~Weinberg and A.~Wu, Phys. Rev. D36 (1987) 2474.
\item
\label{ginsparg}
P.~Ginsparg, Nucl. Phys. B170 (1980) 388.
\end{enumerate}
\section*{Table caption}
Leading infrared contributions to $V_{eff}/T^4$, in units of $m^3/T^3$,
for two- and three-loop vacuum diagrams. Big bubbles correspond to loops
where only zero-frequency modes are considered.
\section*{Figure captions}
\begin{itemize}
\item[Fig.1:]
The temperature-dependent effective potential for $\lambda = 0.1$, $v =
250 \gev$ and some representative values of the temperature $T$. Case (a)
corresponds to the approximation of eq.~(\ref{brute}); case (b)
to that of eqs.~(\ref{veffdaisy}), (\ref{delirium}) and
(\ref{carsol}); case (c) corresponds to the solution of the gap equation
(\ref{gap}), with $\Pi = \Pi^{(a)}$; case (d) corresponds to the solution
(\ref{solphi}) of the gap equation (\ref{gapphi}): when $\mbar(\phi,T)$
becomes purely imaginary, the dash-dotted line gives the real part of
$\veff$. In all cases, the potential is (arbitrarily) normalized to zero
at the origin.
\item[Fig.2:]
One-loop self-energy diagrams.
\item[Fig.3:]
Tadpole (a--d) and vacuum (e) daisy diagrams.
\item[Fig.4:]
Solutions of the gap equation and expansion parameters for the same choice
of parameters as in fig.~1: a) the approximate solution (\ref{carsol});
b) the exact solution of the gap equation (\ref{gap}), with $\Pi = \Pi^{(a)}$;
c) the expansion parameters  $\bar{\alpha}(\phi,T)/(2 \pi)$ (dashed lines),
$\bar{\beta}(\phi,T)/(2 \pi)$ (solid lines) and $\bar{\alpha} \bar{\beta}
\gamma$ (dash-dotted lines) corresponding to case b); d) the solution
(\ref{solphi}) of the gap equation (\ref{gapphi}). In a), b) and d), the
solid lines correspond to $T=490,495 \gev$, the dashed
lines to $T_0 = 500 \gev$, the dash-dotted lines to $T=
505,510 \gev$. In c), for each of the three parameters, the temperature $T$
increases by $5 \gev$ steps going from the right curves to the left ones.
\end{itemize}
\end{document}